\documentstyle[eqsecnum,prb,aps,twocolumn]{revtex}
\input epsf.sty
\def\sech{\mathop {\rm sech} \nolimits}
\begin{document}

\draft

\twocolumn[\hsize\textwidth\columnwidth\hsize\csname
@twocolumnfalse\endcsname

\title{Macroscopic Quantum Tunneling of Ferromagnetic Domain Walls}

\author{Hans-Benjamin Braun$^{1}$, Jordan Kyriakidis$^{2}$, and
    Daniel Loss$^{2}$}

\address{$^1$
Paul Scherrer Institut, CH-5232 Villigen PSI, Switzerland
\\}
\address{$^2$
Department of Physics and Astronomy,
University of Basel, \\ Klingelbergstrasse 82,
4056 Basel, Switzerland}

\date{June 1997}
\maketitle

\begin{abstract}

Quantum tunneling of domain walls out of an impurity potential in a
mesoscopic ferromagnetic sample is investigated.  Using improved
expressions for the domain wall mass and for the pinning potential, we
find that the cross-over temperature between thermal activation and
quantum tunneling is of a different functional form than found
previously.  In materials like Ni or YIG, the crossover temperatures
are around 5 mK. We also find that the WKB exponent is typically two
orders of magnitude larger than current estimates. The sources for
these discrepancies are discussed, and precise estimates for the
transition from three-dimensional to one-dimensional magnetic behavior
of a wire are given. The cross-over temperatures from thermal to
quantum transitions and tunneling rates are calculated for various
materials and sample sizes.

\end{abstract}

]

\section{Introduction}

The possibility of observing quantum mechanical behavior at a
mesoscopic scale has recently attracted much experimental and
theoretical interest. First, there is the fundamental issue of
identifying physical systems possessing many degrees of freedom which
support a collective mode that features typical quantum properties
such as superposition behavior, interference effects, or tunneling
through potential barriers. Well-known examples of such systems are
Josephson junctions which have been extensively studied in the past
(for a review see {\it e.g.}~Leggett\cite{leggett92}).  Recently the
focus has shifted towards low-dimensional magnetic
systems\cite{barbara95} such as single-domain ferromagnets and
antiferromagnets, but also towards nonuniform magnetic structures
exhibiting domain (or Bloch) walls.  In the latter case, one envisages
a domain wall trapped by a magnetic pinning center---as provided, for
example, by an impurity lowering the anisotropy energy locally.  The
domain wall can then escape from this potential well by tunneling
through the energy barrier.  The observability of such tunneling
events basically depends on three conditions which can be stated
qualitatively as follows. First, the tunneling barrier should be
neither too high nor too wide. Second, the effective mass associated
with the tunneling dynamics of the Bloch wall, and hence the number of
spins in the wall, should not be too large. These two conditions are
required in order to have a tunneling rate not too small, so that one
can expect a tunneling event to take place within a reasonable amount
of time (typically on the scale of hours or less).  And third, the
crossover temperature which separates the classical regime of barrier
crossing due to thermal activations from the quantum regime of
tunneling should realistically be in the milli-Kelvin range or above.
Clearly, a more precise formulation of these conditions is essential
since they are of fundamental importance for the interpretation of
recent and future experiments in terms of macroscopic quantum
tunneling.  With this motivation, it is our goal to provide such
quantitative estimates in the following.

Although the idea of domain wall tunneling has first been described
for bulk samples in the seventies,\cite{egami73b,egami73,barbara72} it
was not until the work by Stamp and
collaborators\cite{stamp91,chudnovsky92} that this idea has received
wider attention.  The past few years have seen considerable progress
in sample preparation and has made a detailed experimental study of
the relaxation properties in nanowires possible.\cite{hong96} In such
experiments, the observation of a temperature independent relaxation
(or resistance) below a critical temperature is taken as a strong
indication for quantum tunneling.  Such observed cross-over
temperatures lie in the range of 2 to 5 Kelvin.  Resistivity
measurements at low temperatures require metallic samples.  The
presence of conduction electrons, however, may interfere with the
tunneling process by providing a channel for dissipation (although
large domain walls or low conductivities reduce this undesirable
effect).  Insulating samples overcome this problem, but then
experiments more difficult than resistance measurements are necessary.
Rather than resistivity, magnetization can be measured; a depinned
wall which propagates down the sample will be accompanied by a sudden
change in the magnetization.

Theoretical estimates, based on the same model considered here, have
been given before,\cite{chudnovsky92} but as we shall see the
conclusions reached have been too optimistic. In particular, we find
the functional dependence of the crossover temperature on
experimentally important quantities such as the coercivity and domain
wall mass are quite different from earlier calculations. This result
has already been stated in Ref.~\onlinecite[section VII]{braun96}, but
without any details given. Below we provide these details, thus
supporting earlier claims. The value of the crossover temperature is
of considerable interest for the interpretation of experimental
observations since it is usually taken as a strong indication for the
existence of quantum tunneling if the magnetization switching becomes
temperature independent below this crossover temperature. Also,
earlier estimates\cite{chudnovsky92} predict reasonable tunneling
rates for domain walls containing up to $10^6$ spins, whereas we find
that the number of spins in a flat wall should not exceed $10^4$.

The paper is organized as follows. In Sec.~\ref{secmodel} we present
the model for a ferromagnet. We then discuss the conditions under
which transverse spin waves freeze out and the sample can be
considered one-dimensional.  The domain wall mass is derived from the
well known classical soliton solutions, and the origin of the impurity
potential is discussed. In Sec.~\ref{secdepin} we evaluate the
tunneling rates and cross-over temperatures for a domain wall out of a
pinning potential. Explicit numerical examples are given for various
materials such as YIG, Ni, and in particular the very promising
perovskite and (badly) itinerant ferromagnet
SrRuO$_3$.\cite{klein95,note3} These results are summarized in Tables
I--III. Finally, we compare these results in Sec. \ref{secdisc} with
values given previously in Ref.~\onlinecite{chudnovsky92}.

\section{Model}
\label{secmodel}

We consider an elongated ferromagnetic sample (or ``nanowire'')
as depicted in Figure 1.  We assume that the transverse dimensions
$w$ of the sample are
small enough to ensure that the system behaves effectively
one-dimensional (1D) at the typical temperatures
$T$ of an experiment.  Quantitative estimates for $w$ and $T$ will be
given below.  Now the energy of an effectively 1D
ferromagnet extending along the $x$-axis is given by
\begin{eqnarray}
E[\theta,\phi] &=& {\cal A} \int_{-L/2}^{L/2} dx
\Bigl\{ A[(\partial_x \theta)^2
+ \sin^2 \theta (\partial_x \phi)^2] \nonumber
\\ && + K_e \sin^2 \theta + K_h \sin^2 \theta \sin^2 \phi \Bigr\},
\label{energy}
\end{eqnarray}
where ${\cal A}$ is the cross sectional area of the sample and
the sample length  $L$ is assumed to be much larger than the width
of a domain wall. The magnetization has been expressed  in polar
coordinates, ${\bf M}= M_0 (\sin\theta\cos\phi, \sin\theta \sin \phi,
\cos\theta)$ with
$M_0= g \mu_B s/a^3$ the saturation magnetization and $a$ the lattice
constant.  The three terms in the energy density of Eq.~(\ref{energy})
respectively describe isotropic exchange, easy-axis anisotropy (along
$\hat{z}$) and hard-axis anisotropy (along $\hat{y}$).  The anisotropy
terms are of an effective nature and can contain both crystalline and
demagnetizing (shape-induced) contributions.  A typical example of an
elongated sample is shown in Fig.\ 1 where $K_e=K_{e,{\rm cryst}}$ and
$K_h=K_{h,{\rm cryst}}+ 2\pi M_0^2$.\cite{note}

In the absence of dissipation, the dynamics of the magnetization
is governed by the familiar Landau-Lifshitz equations
\begin{eqnarray}
 \partial_t \phi &=& - {\gamma \over M_0}
 {\delta (E/{\cal A})\over \delta \cos \theta}, \nonumber \\
 \partial_t \cos\theta &=&\phantom{-} {\gamma\over M_0}
 {\delta (E/{\cal A}) \over \delta \phi}.
 \label{LL}
\end{eqnarray}
Here both $\phi$ and $\cos \theta$ depend on $x,t$, the energy
$E$ is given by (\ref{energy}), and
$\gamma = g \mu_B/\hbar$ denotes the gyromagnetic
ratio.

\subsection{1D Regime}

The system exhibits quasi 1D behavior when all
transverse degrees of freedom are frozen out. In order to obtain a
quantitative estimate of this 1D regime  and thus of the validity of the
model  (\ref{energy}), (\ref{LL}), we start from the
three-dimensional (3D) spectrum of excitations around a Bloch
wall, which is  given by\cite{winter62}
\begin{eqnarray}
 \epsilon_{n,\bf k} = &{2a^3\over s} &\sqrt {A {\bf k}_{\perp}^2 +
 n (A k_x^2 + K_e)} \nonumber \\
 &\times & \sqrt{K_h+ A {\bf k}_{\perp}^2 +n (A k_x^2 + K_e)},
 \label{spinwaves}
\end{eqnarray}
and evaluate the corresponding finite size gaps.
Here ${\bf k}_{\perp}=(k_y,k_z)$ is the wave vector of spin waves
running transverse to the sample, and $k_x$ is the wave vector along
the sample. The parameter $n=0,1$ characterizes the type of
excitations. For  $n=1$  one obtains the  spectrum of the
traditional spin-waves, whereas  $n=0$ leads to
the spectrum of Winter (or flexural) wall modes which describe
a curving of the Bloch wall. In the limit of an infinite
sample the spin waves have an anisotropy gap,
$2a^3K_e/s$, while the flexural modes are gapless.  For the {\it finite}
sample widths considered here, however, the transverse spin waves
and in particular the flexural modes acquire an additional
finite size gap. This gap arises since the first excitation in
transverse  direction involves the finite wave vector $k_{\rm
min}=\pi/w$,  where $w$ denotes the maximal transverse dimension of the
sample.   As a consequence of these finite  size gaps, all transverse
excitations of type $n$ around the Bloch wall get frozen out below a
given temperature $T$ for sample widths
$w<w_n$, where $w_n$ follows from Eq.~(\ref{spinwaves})
\begin{equation}
 w_n(T) = { \pi \left[ 2A/(K_h+n K_e) \over  \sqrt{1 +
 (\rho T)^2 -4{n K_e\over K_h +n K_e}}-1\right]^{1/2}}.
 \label{wc}
\end{equation}
Here we have set ${\rho} = s k_B/ (K_h+n K_e) a^3$.
Note that $w_n$ diverges when $k_B T$
approaches $n (2a^3/s) \sqrt{K_e(K_e+K_h)}$ (from above).

Since the minimal energy  $\epsilon_{n=0,\bf k}$
of the flexural wall modes is always smaller than
that of the spin wave modes (at the same wavevectors),
{\it quasi 1D behavior at temperature $T$ is established
for sample widths} $w<w_{n=0}(T)$.  For instance,
we find $w_{n=0}=\pi\sqrt{A/K_h}$ at  a ``freezing''
temperature $k_B T={2\sqrt{2}} K_h a^3/s$, where
$K_h=2\pi M_0^2$ for a slab as shown in Fig.~1. Typical numbers for Ni
(see Table~\ref{material}) are
$w_{n=0}\approx 250 \AA$ and $T\approx 1{\rm K}$.  We note that
these flexural wall modes are frozen out well above the typical
cross-over temperature $T_c$ below which one expects to see quantum
tunneling. This cross-over temperature will be calculated below and is
found to be of the order of $5{\rm mK}$.  For a given sample width
$w$, the transverse spin waves ($n=1$) with $k_x=0,{\bf
k}_{\perp}\neq 0$ are frozen out at even higher temperatures.

\subsection{Soliton solutions and soliton mass}

In  the literature, various differing values for the
wall mass have been used. Therefore we give now a derivation of
the wall mass from the {\it exact}  soliton solutions \cite{walker} of
the equations of motion (\ref{LL}).
These soliton solutions describe Bloch walls traveling at
a constant velocity $v$, and are given by \cite{charge}
\begin{equation}
 \theta_0(x-vt)= 2 \arctan e^{(x-vt)/\tilde\delta}.
 \label{blochwall}
\end{equation}
The soliton velocity $v$ is related to the (constant) azimuthal angle
$\phi_0$ by the expression,
\begin{equation}
 v = \sqrt{{A\over K_e}}  {\gamma K_h\over M_0}
 {\sin 2\phi_0 \over \sqrt{ 1+ \kappa \sin^2\phi_0}},
 \label{vofphi}
\end{equation}
where
\begin{equation}
 \kappa=K_h / K_e.
\end{equation}
We see that at finite velocities the magnetization is tilted out of
the easy $xz$-plane ($\phi_0=0,\pi$) and also that the Bloch wall has a
limiting
velocity (the ``Walker limit'')
\begin{equation} \label{eq-walker}
 v_w=v_0 [\sqrt{1+\kappa} - 1],
\end{equation}
where $v_0=2 \gamma \sqrt{A K_e}/M_0$.

The width of the moving Bloch wall
\begin{equation}
\tilde \delta=\delta[1+ \kappa \sin^2 \phi_0]^{-1/2},
\end{equation}
is contracted relative to the width $\delta=\sqrt{A/K_e}$ of a Bloch
wall at rest. Inserting Eqs.~(\ref{blochwall}) and (\ref{vofphi}) into
(\ref{energy}) we obtain the total energy of a moving Bloch wall
\begin{equation}
 E(\theta_0,\phi_0) = E_0 \sqrt{1+\kappa \sin^2\phi_0},
 \label{ebloch}
\end{equation}
where $E_0=4 {\cal A} \sqrt{A K_e}$ is the energy of a static Bloch
wall.

If the hard-axis anisotropy energy induced by the soliton motion is
much smaller than the easy-axis anisotropy, {\it i.e.}~$\kappa \sin^2
\phi_0 \ll 1$, then $v \ll v_0 \sqrt{\kappa}$, and the energy
in Eq.~(\ref{ebloch}) takes the form
\begin{equation}
 E(\theta_0,\phi_0) = E_0 + {M\over 2} v^2, \label{smallv}
\end{equation}
with the wall mass
\begin{equation}
 M = {\cal A} {M_0^2\over \gamma^2 K_h} \sqrt{{K_e \over A}}
 \label{mass}
\end{equation}
(provided, of course, that $v<v_w$). For a hard-axis of demagnetizing
origin of the form $K_h=2\pi M_0^2$, Eq.~(\ref{mass}) reduces to the
D\"oring expression of the wall mass $M_D=({\cal A}/2\pi \gamma^2)
\sqrt{K_e/A}$.\cite{doring48}  We emphasize that in the presence of an
additional strong crystalline hard-axis anisotropy, $K_h=2\pi M_0^2 +
K_{h,{\rm cryst}}$, with $ K_{h,{\rm cryst}} \gg 2\pi M_0^2$, the wall
mass (\ref{mass}) is substantially smaller than the D\"oring value
(smaller masses lead to higher tunneling rates).  Wall masses that are
up to $10^3$ smaller than the D\"oring value are found in the
orthoferrites.\cite{malozemoff79}

Eq.\ (\ref{smallv}) shows that a moving domain wall behaves
like a particle of mass $M$.  The dynamics of the domain wall
$\theta_0(x-X)$ with $X=X(t)$ can therefore be described by the action
of a free particle of mass $M$,
\begin{equation}
 {\cal S}^{(0)} = \int\! dt\; {M\over 2} \dot X^2.
 \label{sfree}
\end{equation}
For a microscopic derivation of Eq.~(\ref{sfree}) from the
quantum spin action within a coherent spin state path integral
formalism and a collective coordinate technique (and also including
the effects of dissipation via spin waves) we refer the reader to
Refs.~\onlinecite{braun94,braun95a,braun96}.

\subsection{Impurities and pinning potentials}

So far we have focused on an ideal sample with perfect translational
invariance. In realistic samples this invariance is broken by impurities
or modulations of the sample cross section.  We extend now the above
considerations to this situation and discuss the effects of an
external magnetic field.  For simplicity
we treat first a point-like impurity, consisting of a single atom
at ${\bf x}=0$ with easy-axis anisotropy $K_p\neq K_e$.
Such an impurity can be described by changing the anisotropy
constant in Eq.~(\ref{energy}) in the following way,
\begin{equation}
 K_e \to K_e + K_p({\bf x}), \qquad K_p({\bf x})=-V_0 \delta({\bf x}).
\label{imp}
\end{equation}
where $V_0=(K_e-K_p) a^3$, for $\nu$ such impurities we
evidently have $V_0=\nu (K_e-K_p)a^3$. Without loss of generality,
we consider in the sequel the case of attractive impurities,
{\it i.e.}\ $V_0 > 0$.

A uniform external field along the $\hat z$ (easy)-axis is described
by a Zeeman term $-H M_0 \cos\theta$ in the energy density
(\ref{energy}).  Both pinning and external field thus lead to the
additional energy
\begin{equation}
 E'= \int d^3 x \Biggl\{ K_p({\bf x}) \sin^2 \theta
 -H M_0 \cos \theta  \Biggr\}.
\label{sprime}
\end{equation}
The impurity now breaks the translational symmetry transverse to the
sample. We consider here the situation of weak pinning where
the pinning energy is much smaller than the static wall
energy, $V_0/E_0 \ll 1$. In this case deviations from the flat Bloch wall
configuration $(\theta_0,\phi_0)$ can be neglected.
Note that $E_0 = 2 N K_e a^3$ where
$N = 2 {\cal A} \delta/a^3$ is the number of spins in the static wall.
Therefore the weak pinning assumption can be satisfied even in the case of
many impurities as long as the concentration of impurities within the
wall volume is small.

To lowest order in $V_0/E_0$, we can then insert the static soliton
solution $\theta_0(x-X)$ into Eq.~(\ref{sprime}) and obtain the
additional energy \cite{charge}
\begin{equation}
 E' =  V_p(X)  - h X
 \label{eprime}
\end{equation}
with the pinning  potential
\begin{equation}
 V_p(X) = - V_0 \;{\rm sech}^2 \left( {X\over \delta} \right),
\label{V}
\end{equation}
and the force due to the external field
\begin{eqnarray} \label{field}
 h = 2 {\cal A}  M_0  H.
\end{eqnarray}
In (\ref{V}) we have used that $\tilde\delta=\delta(1+{\cal
O}(v/v_0\sqrt{\kappa})^2)$. Note that even a point-like pinning center
of the form (\ref{imp}) creates a shallow potential (\ref{V}) varying
on the length scale of the Bloch wall width $\delta=\sqrt{A/K_e}$.

The pinning potential (\ref{V})  not only holds for
point-like impurities but  also describes pinning
due to {\it variations in the cross-section}
if they extend over length scales $l$ shorter than the domain wall width
$\delta$. To be specific, let us consider a constriction
where the cross sectional area
${\cal A}(x)= {\cal A}- \Delta {\cal A}(x)$
is locally reduced, {\it i.e.} $\Delta {\cal A}(x)$ vanishes for
$|x|> l$.  Let $\Delta v=\int\! dx\,\Delta {\cal A}(x)$ denote
the missing sample volume of the constriction.
The total wall energy is then
\begin{eqnarray}
E'&=& \! \int \! dx \, {\cal A} (x)
\left\{2 K_e  \sin^2 \theta - H M_0  \cos\theta
\right\}   \nonumber \\
&=& - 2 K_e \, \Delta v\; {\rm sech}^2 {X\over \delta} -h X+ {\rm const.}
\label{cross}
\end{eqnarray}
where $\theta=\theta_0(x-X)$.
Thus the effect of the constriction is again described
by Eqs (\ref{eprime}), (\ref{V}) but now with
$V_0=2K_e \Delta v= E_0 \Delta v/ 2 {\cal A} \delta$. The weak-pinning
limit is thus justified as long as the volume
$\Delta v$ is small compared to the volume $2{\cal A} \delta$
occupied by the domain wall. In the second line of Eq. (\ref{cross}),
we have suppressed a small additional
Zeeman term $(h\Delta v/2{\cal A})\tanh(X/\delta)$ which
is an irrelevant constant for large $X$, while for small $X$
it renormalizes $h$ by a factor $1-\Delta v/2{\cal A} \delta$.
However, this renormalization is small in the
weak pinning limit considered here and thus experimentally
not relevant.

In conclusion, we find that the dynamics of a domain wall in
an external field, and in the presence of a point-like impurity (or
a constricted cross section) is described by the action
${\cal S}={\cal S}^{(0)} -\int dt E'$. Explicitly we have
\begin{equation}
 {\cal S}=\int\! d t \Biggl\{ {M\over 2} \dot X^2  - V_p(X)
 + h X  \Biggr\}.
 \label{actiontot}
\end{equation}
with $V_p$ as in (\ref{V}). $V_0$ depends on the
impurity or constriction parameters as defined above.
The dynamics of the Bloch wall is thus seen to be
equivalent to that of a particle of mass
$M$ in a potential $V_p$ under a force $h$.

\section{Depinning via Quantum Tunneling}
\label{secdepin}

In this section we calculate the tunneling probability of a Bloch wall
out of a pinning potential $V(X)$. For the moment, let us consider a
pinning potential of arbitrary shape, as might arise {\it e.g.}~from
the presence of many randomly located impurities. We shall return
below to the specific case of the generic ${\rm sech}^2$ potential.

Interested in tunneling phenomena, we consider the Euclidean
version of the action (\ref{actiontot})
\begin{equation}
 {\cal S}_E = \int_0^{\beta} d\tau \Biggl\{  {M\over 2}
 \left( dX\over d\tau\right)^2 + U(X) \Biggr\},
 \label{se}
\end{equation}
where units have been chosen such that $\hbar=1$.
The potential energy for the wall is given by
\begin{equation}
U(X)= V(X) - h X.
\label{Upot}
\end{equation}
In Eq.~(\ref{se}), $\beta=1/k_B T$, the wall mass $M$ is given by
Eq.~(\ref{mass}), and $V(X)$ is some smooth pinning potential
which, for the present, we keep arbitrary. It is only
assumed that $V(X)<0$ and that it tends to zero for $|X|\to \infty$.
It then follows that $V(X)$ has at least two inflection points $X_i$,
defined by $V''(X_i)=0$.

Let us consider the situation of a vanishing external field where the
wall is pinned at a local minimum $X_{\rm min}$ of $V(X)$. Let $X_i$
be the inflection point closest to the right of $X_{\rm min}$.  Thus
$V'(X_i)>0$ and $V^{(3)}(X_i)<0$.  At small (positive) values of the
external field, the wall is still trapped at $X_{\rm min}$, but as the
field is increased, the potential becomes increasingly tilted and
finally, the metastable state ceases to exist at the coercive force
\begin{equation}
h_c= V'(X_i),
\label{hc}
\end{equation}
where $h_c=2 {\cal A} M_0 H_c>0$, with $H_c$ the classical (zero
temperature) coercivity. In Fig.~\ref{curves} we plot the potential
energy $U(X)$ with the $\sech^2$ pinning potential of
Eq.~(\ref{V}). The three curves shown are for external fields near the
classical coercivity.  It should be kept in mind, however, that the
following analysis is valid for \emph{arbitrary} pinning potentials
(subject to the conditions expounded in the preceding paragraph).

The possibility of quantum tunneling arises when the external field is
close to the classical coercivity, {\it i.e.}
\begin{equation}
0<\epsilon= 1 - H/H_c \ll 1.
\end{equation}
The potential (\ref{Upot}) can then be expanded around the
inflection point $X_i$ of $U(X)$ to yield
\begin{eqnarray}
    U(x) &=& \frac{1}{6} V^{(3)}(X_i) x^3 +
        \left(-\frac{1}{2} \epsilon h_c V^{(3)}(X_i) \right)^{1/2}
        x^2 \nonumber \\
    &=& {27\over 4} V_{\rm max} \left({x\over d}\right)^2
        \left(1-{x\over d} \right).
\label{cubic}
\end{eqnarray}
Several comments are in order regarding these expressions. First, the
third derivative $V^{(3)}(X_i) <0$ in general depends on
the coercivity. Also, we have shifted coordinates so that the minimum is
now $U(0)=0$. In the second line, we have introduced the
tunneling distance $d > 0$, defined by $U(d) = 0$, and the barrier
height $V_{\rm max}$.
These are explicitly given by\cite{note2}
\begin{equation}
V_{\rm max}={2^{5/2}\over 3} { (h_c \epsilon)^{3/2}\over
[-V^{(3)}(X_i)]^{1/2}},
\label{vmax}
\end{equation}
\begin{equation}
d=3 \sqrt{2} {( h_c \epsilon)^{1/2}\over
[-V^{(3)}(X_i)]^{1/2}},
\label{d}
\end{equation}

For external fields close to the coercivity, the Euclidean action
associated with the tunneling of the domain wall is thus given by
\begin{equation}
 {\cal S}_E [x] = \int_0^{\beta} d\tau
\Biggl\{ {M\over 2} \left({d x\over d\tau}\right)^2 + U(x) \Biggr\}.
\end{equation}
with $U(x)$ as in Eq. (\ref{cubic}). This action is rendered
stationary by the ``bounce'' trajectory
\begin{equation}
 x_b(\tau)=d\sech^2 \omega_b \tau
\end{equation}
which runs from $x=0$ to $x=d$ and {\it back} to $x=0$ for $\tau$
increasing from $-\beta/2$ to $\beta/2$ with $\beta \rightarrow
\infty$. The characteristic tunnel frequency is given by
\begin{equation}
\omega_b = (3 / 2)^{3/2} \sqrt{V_{\rm max} / M d^2}.
\label{frequency2}
\end{equation}
Note that $\omega_b$ is half the
harmonic oscillation frequency in the potential minimum
of $U$. The tunneling action
${\cal S}_0 ={\cal S}[x_b]$ can be calculated without
explicit knowledge of the above bounce trajectory, {\it i.e.}
\begin{eqnarray}
{\cal S}_0 &=&2\sqrt{2M}\int_0^d \! dx \; \sqrt{U(x)},\\
&=&{4 \sqrt{6} \over 5} d \sqrt{M V_{\rm max}} =
{18\over 5} {V_{\rm max}\over \omega_b}.
\label{tunnelaction}
\end{eqnarray}
Note that the factor of 2 in the first equation arises because the
escape rate is determined by the action over the whole bounce which
leads from $x=0$ to $x=d$ and back to $x = 0$.

The escape tunneling rate $P$ for the potential $U$ in
Eq.~(\ref{Upot}) has been calculated before in a different
context [see {\it e.g.}, Weiss,\cite{weiss93} p.~109, Eqs.~(8.12) and
(8.16)].  It is explicitly given by the standard WKB expression
\begin{equation}
P= 4 \omega_b \sqrt{15 {\cal S}_0 / 2 \pi}\,
e^{-{\cal S}_0}.
\end{equation}
Typically, quantum tunneling will be observable if the time between
tunneling events, {\it i.e.}~the inverse escape rate $P^{-1}$, does
not exceed a few hours. For a typical attempt frequency $\omega_b$
(given approximately by the exponential prefactor in $P$) of the order
of $10^9 s^{-1}$ this requires that the exponent ${\cal S}_0/\hbar$ be
less than about 30.

For the observation of quantum tunneling it is also important to
ensure that the thermally activated transition rate {\it over} the
barrier, $P_T=\omega_0 \exp[-V_{\rm max}/k_BT]$, does not exceed the
tunneling rate $P$ {\it through} the barrier. This is the case if the
sample temperature $T$ is less than the cross-over temperature $T_c$,
which can be estimated by equating the corresponding transition rates.
By assuming that the prefactors are approximately equal we have
(after reinstating $\hbar$) the relation $k_B T_c = V_{\rm
max}\hbar/{\cal S}_0$, which yields
\begin{eqnarray}
k_B T_c = {5\over 8 d} \sqrt{2V_{\rm max}\over 3M}
={5\over 18} \hbar \omega_b.
\label{crossT}
\end{eqnarray}

In order to obtain further quantitative understanding, we now apply
the above results to the specific case of the generic pinning potential
$V_p(X)=-V_0 \sech^2{X\over \delta}$. With Eqs.~(\ref{hc}),
(\ref{vmax}), and (\ref{d}) we immediately find that
\begin{equation} \label{eq-triad}
h_c={4\over 3^{3/2}}{V_0\over \delta},\ \ \ \ \
V_{\rm max} = {2\sqrt{2}\over 3} h_c \delta \epsilon^{3/2},\ \ \ \ \
d   = 3 \delta \sqrt{{\epsilon \over 2}}.
\end{equation}
The coercive force is thus linked to intrinsic properties of the
pinning potential---the ratio of potential strength $V_0$ and
characteristic length scale $\delta$. Comparison of these expressions
with Eqs.~(\ref{d}) and (\ref{vmax}) shows now explicitly that
$V^{(3)}(X_i)=-4 h_c/\delta^2$ indeed depends on the coercive force.

The tunneling exponent, cross-over temperature, and tunneling frequency
follow from Eqs.~(\ref{tunnelaction}), (\ref{crossT}),
and (\ref{frequency2}) and are given by
\begin{equation} \label{eq-sech}
    {\cal S}_0 = (6/5) \hbar N s \sqrt{H_c / 2 \pi M_0}\,
    (2 \epsilon)^{5/4},
\end{equation} \begin{equation}
    k_B T_c = (5/18) g \mu_B \sqrt{2 \pi M_0 H_c}\,
    (2 \epsilon)^{1/4},
\end{equation} \begin{equation}
    \omega_b = (g \mu_B / \hbar) \sqrt{2 \pi M_0 H_c} \,
    (2\epsilon)^{1/4},
\label{sechaction}
\end{equation}
where $N=2 {\cal A} \delta/ a^3$ is the total number of spins in the
wall and we have assumed a purely shape-induced hard-axis anisotropy,
{\it i.e.}~$K_h= 2\pi M_0^2$ for a slab geometry. Alternative but
equivalent  expressions for the bounce frequency $\omega_b$, the WKB
exponent ${\cal S}_0/\!\hbar$, and the crossover temperature $T_c$ are
listed in Table~\ref{table1}.

To illustrate the above analytical results with concrete numbers we
have collected in Tables \ref{material} and \ref{results} various
values for several ferromagnetic samples of the shape shown in Fig.~1,
namely Yttrium Iron Garnet (YIG), Nickel, the perovskite
SrRuO$_3$,\cite{klein95,note3} and ``large easy-axis'' materials
considered in Ref.~\onlinecite{chudnovsky92}. From Table \ref{results}
it becomes clear that the typical number $N$ of spins one can expect
to tunnel coherently out of a pinning potential within reasonable time
(a few seconds) is of the order of $10^4$ or less, and that the
associated cross-over temperature $T_c$ is typically less than $10{\rm
mK}$. A stark exception to this is SrRuO$_3$, in which our theory
predicts $10^3$ spins can coherently tunnel approximately once every
millisecond with a crossover temperature of 37 milli-Kelvin. It would
therefore be very interesting to look for domain wall tunneling in
this material.

From Eqs.~(\ref{eq-sech})--(\ref{sechaction}) we see that in order
to optimize the observability of quantum tunneling it would be
desirable to have materials that possess both a large coercivity $H_c$
and a large hard axis anisotropy $K_h$ but with the ratio $H_c / \!
\sqrt{K_h}$ being small. Such materials would have a small WKB
exponent ({\it i.e.}\ a high tunneling rate) {\it and} a high
cross-over temperature. There is, however, some leeway by carefully
choosing the experimentally tunable parameters $N$ and $\epsilon$ (see
{\it e.g.}~SrRuO$_3$ in Tables~\ref{material} and \ref{results}).

\section{discussion}
\label{secdisc}

\subsection{Comparison with earlier results}

A discussion of domain walls tunneling out of impurity
potentials has been given by Chudnovsky, Iglesias and Stamp
\cite{chudnovsky92} which we shall henceforth refer to as CIS (for
earlier work in this field see also the references contained in CIS).
Our results presented in Table III for experimentally vital quantities
such as the WKB exponent ${\cal S}_0$ and the cross-over temperature
$T_c$ to the quantum regime differ substantially from the results
given in section VII of CIS. Therefore a comparison of the two
approaches appears necessary.

Before turning to the most crucial difference between the two
approaches---the functional dependence of the pinning potential on
the coercivity---let us first remark that the mass $M^{\rm CIS}$
used in CIS differs from the soliton mass $M$ given in
Eq.~(\ref{mass}). In fact,
\begin{equation}
M^{\rm CIS}=  M {\kappa\over (\sqrt{1+\kappa}-1)^2}.
\end{equation}
The two masses agree only in the limit of large hard-axis anisotropy,
{\it i.e.}~$\kappa\equiv K_h / K_e \gg 1$. However, in the
experimentally important limit of $\kappa \ll 1$ we have $M^{\rm
CIS}/M=4/\kappa\gg 1$. Thus, we would expect our action to be smaller
and the cross-over temperature larger than the CIS results.

However, this tendency is more than compensated by the CIS assumption
that the pinning potential width $\bar w$ is independent of the
coercivity. For a field close to the coercivity, CIS find (their
Eq.~(25)) for the total pinning potential of arbitrary shape
\begin{equation}
A_w  U(Z) = \sigma_0 A_w \left\{
{\sqrt{2 h_c^{\rm CIS}\epsilon \ } \over 2 \bar w}
\left({Z\over\delta} \right)^2 - {1\over 6 \bar w^2}
\left({Z\over\delta} \right)^3 \right\},
\label{U}
\end{equation}
where $h_c^{\rm CIS}=\delta h_c / \! E_0$, $\sigma_0 A_w = E_0$, and
$A_w = \cal{A}$. $\bar w$ is a parameter which is assumed to be
independent of $h_c$ and is set equal to unity in section VII of CIS.
However, comparison with our Eq. (\ref{cubic}) reveals that the parameter
$\bar w^2 = -\sigma_0 A_w/V^{(3)}(X_i)\delta^3$ is  not arbitrary but
depends on the details of the  pinning potential.

In particular, for the  generic ${\rm sech}^2$ pinning potential
(\ref{V}) it follows that $\bar w$ is coercivity-dependent:
\begin{equation}
\bar w = {1 \over 2 \sqrt{h_c^{\rm CIS}}}.
\label{w}
\end{equation}
Since experimentally $h^{\rm CIS}_c=H_c/H_a$ (where $H_a=2 K_e/M_0$)
lies typically in the range $10^{-2}$--$10^{-5}$,
this implies a value of
$\bar w$ in the range 5--100.

By assuming $\bar w \simeq 1$ in their final section, CIS have thus
{\it a priori} fixed the depth of the pinning potential to the
extremely large value of a third of the total wall energy,
$V_0=E_0/3$. This  situation corresponds to a region of length $4
\delta/3$ extending across the entire cross sectional area being
replaced with magnetic ions of vanishing anisotropy.

If we insert our expressions for $M$ and $\bar w$ into Eq.~(27)
of CIS, then their tunneling action agrees with ours (up to a minor
error of a missing factor of $2^{1/4}$ in Eq.~(27) of CIS). With these
substitutions we also find agreement between our $T_c$ and Eq. (44) of
CIS (where again a factor $2^{-1/4}$ is missing).

Our results strikingly differ from CIS when it comes to the explicit
computation of experimentally important quantities. In Eqs.~(88) and
(87) of CIS, numerical factors $2^{1/4} 48/5 \simeq 11 $ and
$5\sqrt{2}/2^{1/4} 36\simeq 1/6$ respectively are suppressed compared
to CIS Eqs.~(27) and (44) for ${\cal S}_0$ and $T_c$. Together with
their assumption that $\bar w \simeq 1$,  this leads  (for a material
with $h_c^{\rm CIS}=10^{-3}$) to an underestimation of the WKB
exponent ${\cal S}_0$ by a factor of $700$ for a planar domain wall
tunneling through the potential of a single defect. At the same time
the cross-over temperature $T_c$ between quantum tunneling and thermal
depinning is overestimated by a factor of $24$.

Finally, we mention that while our cross-over temperatures and WKB
exponents differ substantially from the values presented in CIS, we
find the same scaling of $T_c$ and ${\cal S}_0$  with respect to the
reduced field $\epsilon$.

\subsection{A mechanism for increasing ${\bf T_c}$}

Explicit expressions for various materials such as Ni, Fe, YIG, and
${\rm Sr}{\rm Ru}{\rm O}_3$ are presented in Table \ref{results}. It
is seen that the transition temperatures are in the ${\rm mK}$ range.
A notable exception is ${\rm Sr}{\rm Ru}{\rm O}_3$, whose small domain
wall width leads to a narrow potential well and thus to a considerably
higher transition temperature---around $40 {\rm mK}$.

We must mention, however, that our simple analysis here does not rule
out discernible tunneling of larger walls at higher temperatures.
Recent experiments on domain wall dynamics\cite{hong96} have been
interpreted as evidence of tunneling. This evidence primarily stems
from the occurrence of temperature independent phenomena below a
cross-over temperature of a few Kelvin---{\it three orders of
magnitude larger than our estimates here}. Rather than comment on
these experiments, we will instead discuss a plausible mechanism which
may raise $T_c$ and/or decrease ${\cal S}_0$.

Table \ref{results} shows that the tunneling rate becomes appreciable
if the tunneling distance is smaller than $10 \AA$. Thus, variations
of the pinning potential $V(X)$ on this length scale could
dramatically affect the tunneling behavior. Such variations do not
occur for a random superposition of pinning potentials of the
$\sech^2$ type. However, under certain circumstances the underlying
crystal lattice can provide such a modulation, in particular if the
wall width is only a few lattice constants.

To get some quantitative ideas about the consequences of a modulation
with the period of the lattice constant, let us add the term $V_{\rm
per}(X) = V_1 \sin ( 2 \pi X / a + \zeta)$ to the pinning potential in
Eq.~(\ref{Upot}) Here, $\zeta$ is a phase which we conveniently choose
as $\zeta = -2\pi X_i / a$, where $X_i$ is the inflection point of the
$\sech^2$ potential, {\it i.e.}~$\sech^2 X_i/\delta=2/3$. Thus $X_i$
remains the inflection point of $V+V_{\rm per}$. This new potential
has associated with it a coercivity which is a factor of $(1+\mu)$
larger than the coercivity in Eq.~(\ref{eq-triad}), {\it i.e.}~$h_c =
h_c^{\rm old}(1+ \mu)$, where
\begin{equation} \label{eq-coerc.new}
    \mu = \frac{\sqrt{27}\, \pi}{2} \frac{V_1 \delta}{V_0 a}.
\end{equation}
In the vicinity of $X_i$, we can carry out the same expansion as
outlined above to again obtain the cubic potential given by
\begin{equation} \label{eq-vnew}
    V(x) = \sqrt{2 \epsilon} \frac{h_c}{\delta} \lambda x^2 -
    \frac{2 h_c}{3 \delta^2} \lambda^2 x^3.
\end{equation}
The new parameter $\lambda$, which equals unity for $V_1 = 0$, is
given by
\begin{equation}
    \lambda = \left( \frac{1 + r^2 \mu}{1 + \mu} \right)^{1/2},
\end{equation}
where $r = \pi \delta / a$. In Eq.~(\ref{eq-vnew}) we have also
redefined the reduced field $\epsilon$ to reflect the increased
coercivity, {\it i.e.}
\begin{equation} \label{eq-epsnew}
    \epsilon = 1 - \frac{h}{h_c}
    = \frac{ \epsilon^{\rm old} + \mu }{1 + \mu}.
\end{equation}
Note that we now remain in the tunneling regime ({\it i.e.}~$\epsilon
> 0$) even if $\epsilon^{\rm old}$ becomes negative.
Eq.~(\ref{eq-vnew}) gives rise to a new tunneling distance $d$ and
also a new barrier height $V_{\rm max}$, which results in a larger
crossover temperature $T_c$ and a reduced WKB exponent ${\cal S}_0$.
These new expressions are explicitly given by
\begin{equation}
    T_c = T_c^{\rm old} \sqrt{(1 + \mu) \lambda\, }
\end{equation}
\begin{equation}
    {\cal S}_0 = {\cal S}_0^{\rm old}
    \sqrt{ (1 + \mu) /\! \lambda^3\, },
\end{equation}
where $T_c^{\rm old}$ and ${\cal S}_0^{\rm old}$ contain the {\it new}
definition of $\epsilon$, Eq.~(\ref{eq-epsnew}), since it is now
$\epsilon$ (and not $\epsilon^{\rm old}$) which is the experimentally
tunable parameter controlling the barrier height for tunneling. To
obtain a numerical estimate of this effect, we need to estimate the
magnitude $V_1$ of the periodic piece relative to the impurity
strength $V_0$. Let us take, for example, $V_1 = 10^{-1} V_0$ and a
domain wall of ten lattice constants, $\delta = 10 a$. In this case,
$T_c \approx 10 T_c^{\rm old}$ and ${\cal S}_0 \approx 10^{-2} {\cal
S}_0^{\rm old}$. The estimates here for the WKB exponent must be taken
with caution because these are estimates for the tunneling through
{\em only one} of the (periodic) barriers. Due to the shape of the
impurity potential, one should expect that there are $\sim \delta/a$
such barriers to tunnel through before the wall is free. Assuming
incoherent sequential tunneling, ${\cal S}_0$ will effectively
increase by a factor of about 10 for the estimates just given, {\it
i.e.}~${\cal S}_0^{\rm eff} \approx 10 {\cal S}_0$. Nevertheless, this
very simple argument shows that observation of tunneling of larger
walls at higher temperatures is not necessarily ruled out.

The case of a periodic potential superimposed on the impurity
potential is interesting from another perspective. For $\epsilon^{\rm
old} = 0$ (but $\epsilon > 0$), the soliton originally pinned at the
impurity now sees an effectively flat periodic potential through which
it may tunnel {\it coherently}. Since the soliton is a particle-like
excitation, we basically have here the physics of a particle in a
periodic potential, and thus the soliton can form Bloch-like states of
definite momentum.\cite{braun96} Increasing the field still further
induces a force on the soliton and the possibility arises of Bloch
oscillations {\it of a magnetic soliton}. This idea was first laid out
semiclassically in Ref.~\onlinecite{braun94} and essentially the same
physics holds down to the extreme quantum case of
spin-1/2.\cite{note4}

Finally, we briefly mention that thermally assisted tunneling may also
raise the effective $T_c$. If the pinning potential contains some
internal level structure, then the wall may be thermally excited to
some higher level, and only then tunnel out of the pinning site. A
more detailed analysis of this problem ({\it e.g.}~along the lines
discussed in the context of MQT in SQUIDS, see Weiss\cite{weiss93}) is
required, however, before concrete statements concerning $T_c$ can be
made.

\section{Conclusions}

We have given in this paper a detailed derivation of the tunneling
problem of a planar Bloch wall out of a pinning potential. We have
focused exclusively on a quasi-one-dimensional ferromagnet with
biaxial anisotropy, and have given estimates on when a system can be
considered quasi-one-dimensional. In particular, the flexural
spin-wave modes, while gapless in infinite systems, acquire a gap for
the finite geometries shown in Figure~1. If the sample temperature is
below this energy gap, then the flexural modes cannot be excited and
can hence be neglected. For cross-sectional areas on the order of
$10^4\AA^2$, this energy gap is much larger than the crossover
temperature at which quantum and thermal transitions are equal.

We have modeled the pinning center as an impurity which decreases the
easy-axis anisotropy at a single point in space. Even such a
point-like impurity produces a shallow pinning potential which varies
on the length scale of the Bloch wall width. We have related both the
height and width of this potential to a coercivity. Detectable
tunneling can only occur if the external field is very close to this
coercivity, {\it i.e.}\ we must have $\epsilon \equiv 1 - H/H_c$ on
the order of $10^{-2}$ or $10^{-3}$. This is an important number to
determine experimentally. For example, if an experiment has $\epsilon
\leq 0$, then one would also observe temperature independent
depinning, but of course this cannot be ascribed to quantum
tunneling--- it is trivially due to the vanishing of the barrier height.

Within an instanton approach, the WKB exponent, bounce frequency,
tunneling rate, and crossover temperature have been calculated and
different analytical forms for these quantities can be found in
Table~\ref{table1}. We have also given estimates of these quantities
for specific materials. The material parameters can be found in
Table~\ref{material}, and the estimates in Table~\ref{results}. In
particular, the perovskite SrRuO$_3$ seems a promising candidate and
we hope this work can motivate some experimental studies into this
material.

We have compared our results with previous work\cite{chudnovsky92} and
have found our calculations to predict lower crossover temperatures
and a lower maximum number of spins which can coherently tunnel out of
a pinning potential.

Finally, we have briefly discussed how the effects of a periodic
potential, may lead to a larger crossover temperature and a smaller
WKB exponent.

\section{Acknowledgments}

We wish to acknowledge the hospitality of the ITP, Santa Barbara, where
part of this work has been performed.  This research was supported in
part by the Swiss NSF, the US NSF under Grant No.~PHY94-07194, and by
the NSERC of Canada.

\newpage

\begin{figure}
\epsfxsize=8cm
\leavevmode
\epsfclipon
\epsfbox{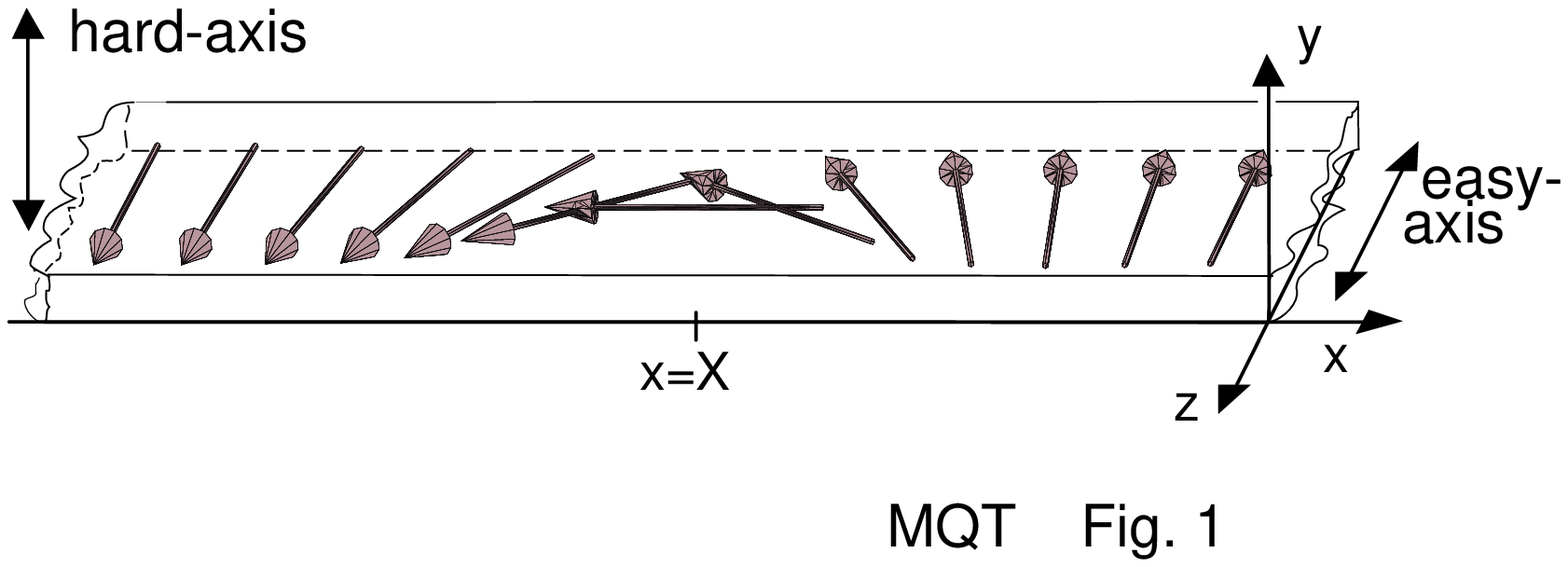}
\epsfclipoff
\caption{Shown is a Bloch wall configuration in a thin long slab, {\it
i.e.}\ $\theta_0(x-X)$, and $\phi_0=\pi$.}
\label{sample}
\end{figure}

\begin{figure}
\epsfxsize=8cm
\leavevmode
\epsfclipon
\epsfbox{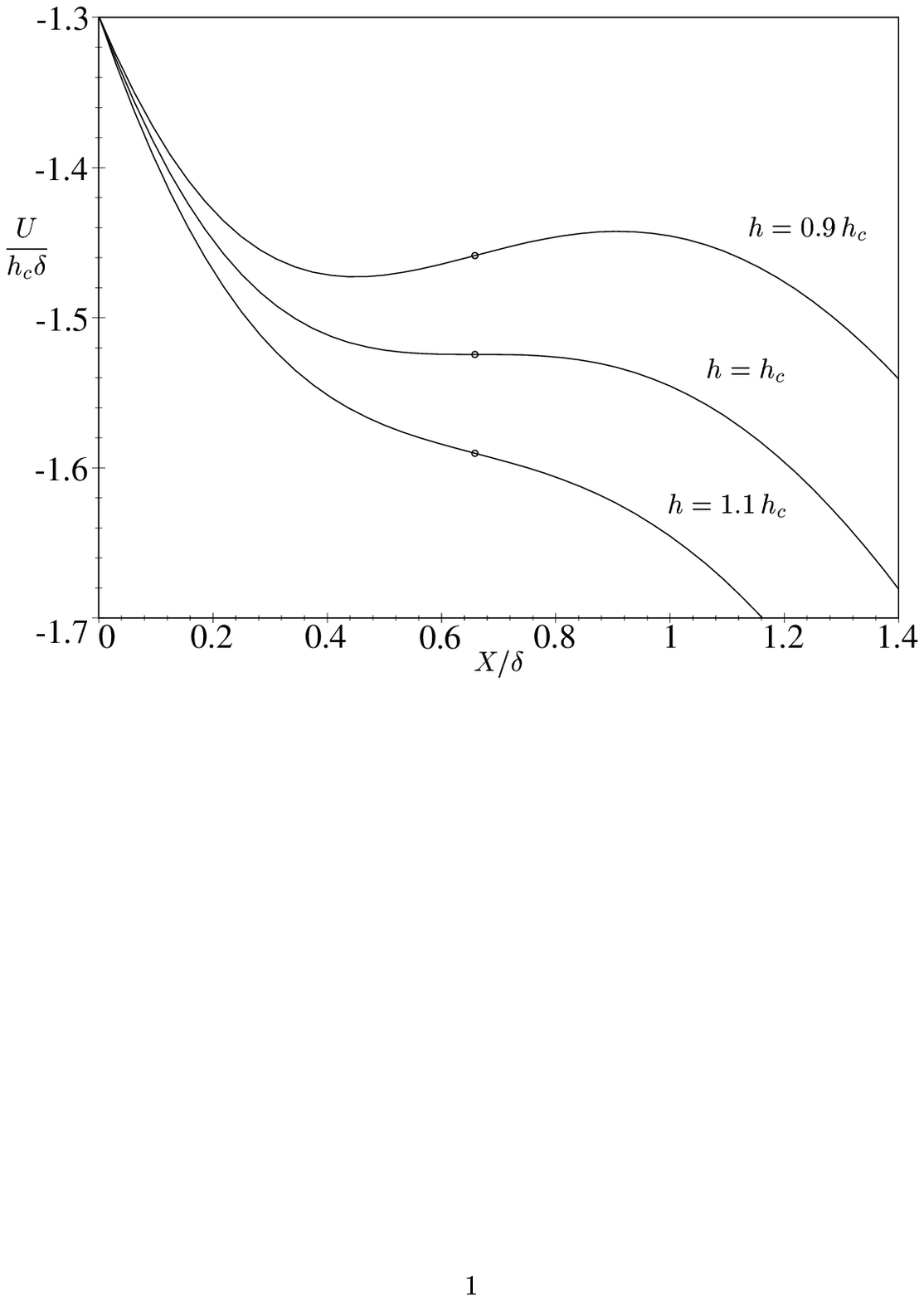}
\epsfclipoff
\caption{We plot the potential energy $U(X)$ of Eq.~(\ref{Upot}) for
the $\sech^2$ pinning potential of Eq.~(\ref{V}). For this pinning
potential, the coercivity $h_c$ equals $(4 / \sqrt{27}) V_0 / \delta$.
The three curves show the potential energy for the external field $h$
slightly above, right at, and slightly below the coercivity field
$h_c$.  By expanding about the inflection point (shown by small
circles in the figure), these curves are very well approximated by a
cubic potential, as discussed in the text.}
\label{curves}
\end{figure}

\onecolumn
\widetext

\begin{table}
\caption{Summary of equivalent expressions for bounce
frequency $\omega_b$, WKB exponent ${\protect\cal S}_0/\hbar$, and cross-over
temperature $T_c$. $N=2\delta {\cal A}/ a^3$
denotes the number of spins in the wall. In the last column it has been
assumed that
$K_h=2 \pi M_0^2$, {\protect\it i.e.}~the sample has the slab geometry of
Fig.~1 and there is no crystalline hard-axis anisotropy, $K_{h,
{\protect\rm cryst}}=0$.}

\begin{tabular}{l|c|c|c|c}
$\omega_b$
     & $ \left(3\over 2 \right)^{3/2} {1\over d} \sqrt{{V_{\rm max}\over M}} $
     & $ \sqrt{{h_c\over M \delta}} \left({\epsilon\over 2}\right)^{1/4} $
     & $\gamma \sqrt{2H_c K_h\over M_0}\left(\epsilon\over 2\right)^{1/4} $
     & $2^{3/4}{g\mu_B\over \hbar}\sqrt{H_cM_0\pi} \epsilon^{1/4} $
\\
\hline
${\cal S}_0$
     & $\sqrt{{3\over 2}} {8\over 5} d\sqrt{M V_{\rm max} }$
     & $2^{3/4} {12\over 5}  \sqrt{M h_c} \delta^{3/2} \epsilon^{5/4}$
     &  $2^{3/4} {6\over 5}\hbar N s\sqrt{{2H_c M_0\over
K_h}}{\epsilon}^{5/4}$
     & $2^{3/4} {6\over 5}\hbar N s \sqrt{{H_c\over \pi M_0}}{\epsilon}^{5/4}$
\\
\hline
$T_c$
     & $ \sqrt{{2\over 3}} {5\over 8} {\hbar\over k_B} {1\over d}
       \sqrt{{V_{\rm max}\over M}}$
     & $ 2^{3/4} {5\over 36} {\hbar\over k_B}\sqrt{{h_c\over M \delta}}
\epsilon^{1/4} $
     & $ {5\over 18} {g \mu_B\over k_B} \sqrt{{K_h H_c\over M_0}}
(2\epsilon)^{1/4} $
     & $ 2^{3/4}{5\over 18} {g \mu_B\over k_B} \sqrt{\pi H_c M_0}
\epsilon^{1/4} $
 \end{tabular}
 \label{table1}
 \end{table}

\begin{table}
\caption{ Saturation magnetization $M_0$, shape anisotropy $K_h=2\pi M_0^2$
for a thin film, easy-axis anisotropy constant $K_e$, exchange $A$,
wall width $\delta=\sqrt{A/K_e}$, wall mass $M$, and coercivity $H_c$ for
various
materials.}

\begin{tabular}{l|c|c|c|c|c|c|c}
 &$M_0 $  &$K_h$ &$K_e$ &$A$  &$\delta$  &$M/{\rm area}$  &$H_c$
\\
 &$[{\rm Oe}]$
      &$[10^5{{\rm erg}\over {\rm  cm}^3}]$
          & $ [10^5{{\rm erg}\over {\rm cm}^3}]$
             &$ [10^{-6}{{\rm erg}\over {\rm cm}}]$
                &$ [\AA]$
   &$[10^{-10}{{\rm g}\over{\rm cm}^2}]$
             &$[{\rm Oe}]$
\\   \hline
 YIG
    & 196 \tablenote{Ref.~\onlinecite{odell81}, p.~65}
        & $2.4$
              & $0.25$ \tablenote{Ref.~\onlinecite{tebble69}, p.~313}
                   & $ 0.43 $ \tablenotemark[1]
                        & 414
                           & $1.2$
                              &$10$
  \\
Ni
    &508 \tablenote{Ref.~\onlinecite{bozorth51}, p.~270}
        & $16$
              & $8$ \tablenote{Ref.~\onlinecite{bozorth51}, p.~569}
                   & $1 $
                        & $ 112 $
                             & $4.6 $
                        &$100$
  \\
  large
   &200
        & $2.5 $
              & $100 $
                   & $1$
                        & $ 32 $
                             & $16 $
                   &$10$
  \\
  $K_e$
\tablenote{Example given in Ref.\protect{\onlinecite{chudnovsky92}},
parameters
taken from there except for $A$ which has been replaced by the most
common value} &&&&&&
\\ \hline
SrRuO$_3$ \tablenote{Ref.~\onlinecite{klein95}}
 &159 & 1.6 & 20 &0.023 & 11 & 48 & $10^4$
 \end{tabular}
\label{material}
\end{table}

\begin{table}
\caption{ Cross sectional area ${\cal A}$, number of spins $N$ in the wall,
$\epsilon$, tunneling distance $d$, cross-over temperature $T_c$, WKB
exponent ${\cal S}_0/\hbar$, oscillation frequency $\omega$, and
inverse tunneling rate $P^{-1}$ for various materials.}

\begin{tabular}{l|c|c|c|c|c|c|c}
                      & ${\cal A}$
                            &$\epsilon$
                                       & $d$
                                             & $T_c$
                                                   & ${\cal S}_0/\hbar$
                                                   & $\omega$
                                                   & $P^{-1}$
\\
                        & $ [\AA^2]$
                              &
                                       & $ [\AA]$
                                             & $[{\rm mK}]$
                                                   &
                                                    &$[10^9\cdot s^{-1}]$
                                                    &$[s]$
\\  \hline
              YIG
                      & $50 \times 200$
                            &$10^{-1}$
                                       & 280
                                             & 3
                                                    & 1268
                                                   &$2.6$
                                                             &$\infty$ \\
                      & $[N=3.4\cdot 10^4]$
                       &  $10^{-2}$
                                        & 88
                                               &1.6

                                    & 71
                                                  &$1.5$
                                                   &$2\cdot 10^{20}$ \\

                     &&  $5.7\cdot10^{-3}$
                                        &66
                                               &1.4
                                                &31.1
                                                 &$1.3$
                                                   &1433 \\

                     &&  $10^{-4}$
                                        &8.8
                                               &0.5

                          & 0.2
                                          &$0.47$
                                                   &$2\cdot10^{-9}$\\
\hline
 Ni
                     & $50 \times 200$
                           &$10^{-1}$
                                        & 75
                                               & 14
                                                   & 1740
                                          &$13$
                                                   &$\infty$ \\
                      &$[N=2.4\cdot 10^4]$
                     &$10^{-2}$
                                        & 23
                                               &8

                          &  98
                                          &$7.6$
                                                   &$10^{31}$\\

                     &&  $3.9\cdot10^{-3}$
                                        &15
                                               &6.3
                                                &31.1
                                                 &$6$
                                                   &310 \\

                     &&  $10^{-4}$
                                        & 2.4
                                               & 2.5

                          & 0.3
                                          &$2.4$
                                                   &$3\cdot10^{-10}$ \\
\hline

large $K_e$
                     & $50 \times 200$
                            &$10^{-1}$
                                       & 21
                                             & 3
                                                   & 98
                                          &$2.6$
                                                   &$5\cdot10^{31}$  \\
                     &$[N=8.0 \times 10^4]$
                            &$3.6\cdot10^{-2}$
                                       & 13
                                             & 2.1
                                                   & 31.1
                                          &$2$
                                                   &$931$  \\
                     &&$10^{-2}$
                                        & 7
                                               & 1.6

                          &  5.5
                                          &$1.5$
                                                   &$2\cdot10^{-8}$   \\

                     &&  $10^{-4}$
                                        & 0.7
                                               & 0.5

                          & 2$\cdot
10^{-2}$
                                          &$0.47$
                                                   &$5\cdot10^{-9}$ \\
\hline

SrRuO$_3$
                     & $50 \times 200$
                            &$10^{-1}$
                                       & 7.4
                                             & 79
                                                   & 873
                                          &$37$
                                                   &$\infty$  \\
                     &$[N=3.4 \times 10^3]$
                            &$10^{-2}$
                                       & $2.3$
                                             & 44
                                                   & 49
                                          &21
                                                   &$10^9$  \\
                     &&$5 \times 10^{-3}$
                                        & 1.7
                                               & 37

                          & 21
                                          &$18$
                                                   &$10^{-3}$   \\

                     &&  $10^{-3}$
                                        & 0.74
                                               & 25

                          & 2.8
                                          &12
                                                   &$10^{-10}$
\end{tabular}
\label{results}
\end{table}


\begin{thebibliography}{10}
\bibitem{charge} Here we have fixed boundary conditions
to $Q={1\over 2}(\cos\theta(-\infty)-\cos\theta(+\infty))=1$.
Eq.~(\ref{LL}) also admits solutions with $Q=-1$ which are
obtained from (\ref{blochwall}) by the parity transformation
$x \to -x$.  The coupling to an external field along the $\hat z$-axis
in (\ref{eprime}) is then generalized to $-hQX$, thus characterizing
$Q$ as the (topological) ``magnetic charge'' of the soliton.
\bibitem{note2}
It is interesting to note that the ratio $V_{\rm max} / d = 4 h_c
\epsilon / 9$ only depends on the coercivity and is otherwise
independent of the details of the pinning potential $V(X)$.
\end{thebibliography}
\end{document}